# Porosification of the dielectric layer in alternating-current electroluminescence devices enables high-resolution tactile imaging

Mingwei Gu, Licheng Li, Zeyuan Zhou, Ying Yang, Zixiang Nie, Xingyu Ma, Junshuai Chen, Lin Zheng, Jinhui Gu, Jia-Sen Xie, Ziling Wei, Mengxing Huang, Jin Ge*

MOE Laboratory of Bioinorganic and Synthetic Chemistry, State Key Laboratory of Optoelectronic Materials and Technologies, GBRCE for Functional Molecular Engineering, LIFM, IGCME, School of Chemistry, Sun Yat-Sen University, Guangzhou 510006, China

*Corresponding author: gejin@mail.sysu.edu.cn;

.

## Abstract

Vision-based tactile sensing enables high-resolution contact perception, yet bulky optical architectures consume valuable space and constrain the functional design of dexterous robotic hands. Here we introduce a self-emissive tactile imaging strategy based on alternating-current electroluminescence (ACEL) devices, in which a compressible porous dielectric encodes contact pressure into spatial emission patterns without external illumination or discrete sensing pixels. We identify a nonlinear capacitance-dependent emission window that provides a basis for tunable pressure-to-light conversion. Pore-size engineering further suppresses local emission deficiencies, reducing spatial intensity variation by nearly ninefold and enabling faithful imaging of contact features down to 100 μm. The continuous electroluminescent interface also reproduces complex contact patterns without computational reconstruction of electrical fields. Direct lamination of the ACEL device onto a complementary metal–oxide–semiconductor (CMOS) image sensor is expected to enable compact contact imaging in an assembly less than 5 mm thick, substantially thinner than conventional vision-based tactile sensors. This approach could reduce the space required for high-resolution tactile imaging, offering greater design freedom for dexterous robotic hands.

## 1. Introduction

The growing demand for dexterous manipulation in embodied robotics calls for tactile sensors capable of resolving complex physical interactions with the environment[1-2]. Beyond measuring contact forces, tactile sensing provides spatial information on pressure distributions, surface features, and dynamic contact states, enabling fine object perception and adaptive manipulation[3-5]. Advances in materials and device engineering have enabled diverse tactile sensing mechanisms, including piezoresistive, capacitive, piezoelectric, triboelectric, magnetic, and optical approaches[6-8]. These technologies have advanced tactile perception from discrete force measurements to high-density pressure mapping, with improved sensitivity, response speed, and mechanical compliance[9-11]. However, achieving high spatial resolution, uniform response, and faithful reconstruction of complex contact patterns within a continuous, flexible sensing interface remains challenging[12-13].

High-resolution tactile imaging has been pursued through dense sensor arrays, artificial intelligence (AI)-assisted super-resolution, and physical field reconstruction[10, 14-16]. Dense taxel arrays improve spatial resolution through sensor miniaturization[9, 17], but require complex electrical interconnections and addressing schemes, with crosstalk further limiting scalability[18]. Alternatively, AI-assisted approaches reconstruct fine pressure distributions from sparse measurements[10, 19], yet their accuracy depends on training data and model generalizability[20]. Electrical impedance tomography (EIT), recover continuous pressure distributions from boundary measurements[15-16], but suffer from noise sensitivity and the ill-posed nature of inverse reconstruction[21]. These approaches rely on dense sensor integration or computational reconstruction, highlighting the need for more direct strategies to capture spatial pressure distributions without compromising resolution or increasing system complexity.

Vision-based tactile sensing captures dense contact information optically, avoiding individually addressed electrical sensing elements[22-23]. Current approaches, including marker-based[22], reflective-layer-based[24], and mechanochromic sensing systems[25-26] (**Figure 1A)**, enable high-resolution contact imaging but generally require external illumination and space-consuming optical architectures[23, 25]. Mechanoluminescent materials eliminate the need for external illumination [27-29], but stress-change-dependent emission can limit sustained static-pressure imaging[11]. Alternatively, alternating-current electroluminescence (ACEL) offers electrically driven light emission from continuous, flexible interfaces (Figure 1B)[30-32]. Its architectures have expanded from conventional stacked structures to coplanar electrodes[33] and coaxial fibers[34-35], supporting applications in flexible displays[32], smart textiles[36], and interactive sensing systems[37], The sensitivity of ACEL emission to local impedance and electric fields provides a mechanism for directly encoding contact pressure into special light-intensity distributions[12, 38-39]. Existing strategies exploit resistance modulation[12], interfacial microstructures[11], and modulus-gradient engineering[40] to achieve contact pressure imaging. Yet these approaches offer limited control over pressure-dependent light emission or require complex microfabrication, compromising reproducibility and spatial imaging fidelity. Achieving broadly tunable, high-resolution pressure imaging through a scalable, compact ACEL architecture remains challenging.

Here we introduce a porous-dielectric strategy for high-resolution tactile imaging through pressure-modulated alternating-current electroluminescence (ACEL). By exploiting the nonlinear dependence of light emission on series capacitance, we identify an effective modulation window that enables tunable pressure-to-light conversion. A compressible porous dielectric translates local pressure into spatial variations in emission intensity through deformation-induced redistribution of the electric field. Pore-size engineering further improves emission uniformity and spatial fidelity, enabling direct imaging of contact features down to approximately 100 μm without discrete sensing elements or complex inverse reconstruction. Experimental characterization and finite-element simulations reveal the underlying coupling between dielectric deformation, electric-field redistribution, and electroluminescent response. The self-emissive ACEL interface allows direct lamination onto a complementary metal–oxide–semiconductor (CMOS) image sensor without external bulky illumination system, potentially enabling tactile imaging in an assembly less than 5 mm thick. This approach offers a tunable and potentially scalable route towards high-resolution tactile imaging while reducing the space demands imposed on dexterous robotic hand design.

## 2. Results and Discussion

### 2.1 Effective capacitance modulation window for ACEL emission

The emission intensity of alternating-current electroluminescent (ACEL) devices depends strongly on the electric field across the emissive layer, which can be regulated through capacitive voltage division. To establish a quantitative relationship between capacitance and light emission, we introduced an external variable capacitor in series with an ACEL device (Figure 1C). An equivalent circuit comprising the emissive-layer resistance ($R_{EL}$) and capacitance ($C_{EL}$), series resistance ($R_S$), and external variable capacitance ( $C_{ext}$) was established (Figure 1C). The voltage across the emissive layer and the corresponding complex impedances are described by Equations (1) and (2), respectively, while the average electric-field amplitude is given by Equation (3). Increasing the external capacitance reduces its capacitive reactance, increasing the voltage across the emissive layer until the voltage division approaches a limiting value.

$$V_{EL} = V_{AC} \frac{Z_{EL}}{R_S + Z_{EL} + Z_{ext}} \tag{1}$$

$$Z_{EL} = \left(\frac{1}{R_{EL}} + j\omega C_{EL}\right)^{-1}, \; Z_{ext} = \frac{1}{j\omega C_{ext}} \tag{2}$$

$$E_{EL,nom} = \frac{|V_{AC}|}{d_{EL}} \left| \frac{Z_{EL}}{R_S + Z_{EL} + Z_{ext}} \right| \tag{3}$$

Here, $V_{AC}$ and $V_{EL}$ denote the complex voltage phasors of the applied field and emissive layer, respectively; $Z_{EL}$ and $Z_{ext}$ represent the impedances of the emissive layer and the external variable capacitor; $R_{EL}$ and $C_{EL}$ are the equivalent resistance and capacitance of the emissive layer; $R_S$ is the series resistance; $C_{ext}$ is the external capacitance; $\omega = 2\pi f$ is the angular frequency, where $f$ is the driving frequency; and $j$ is the imaginary unit.

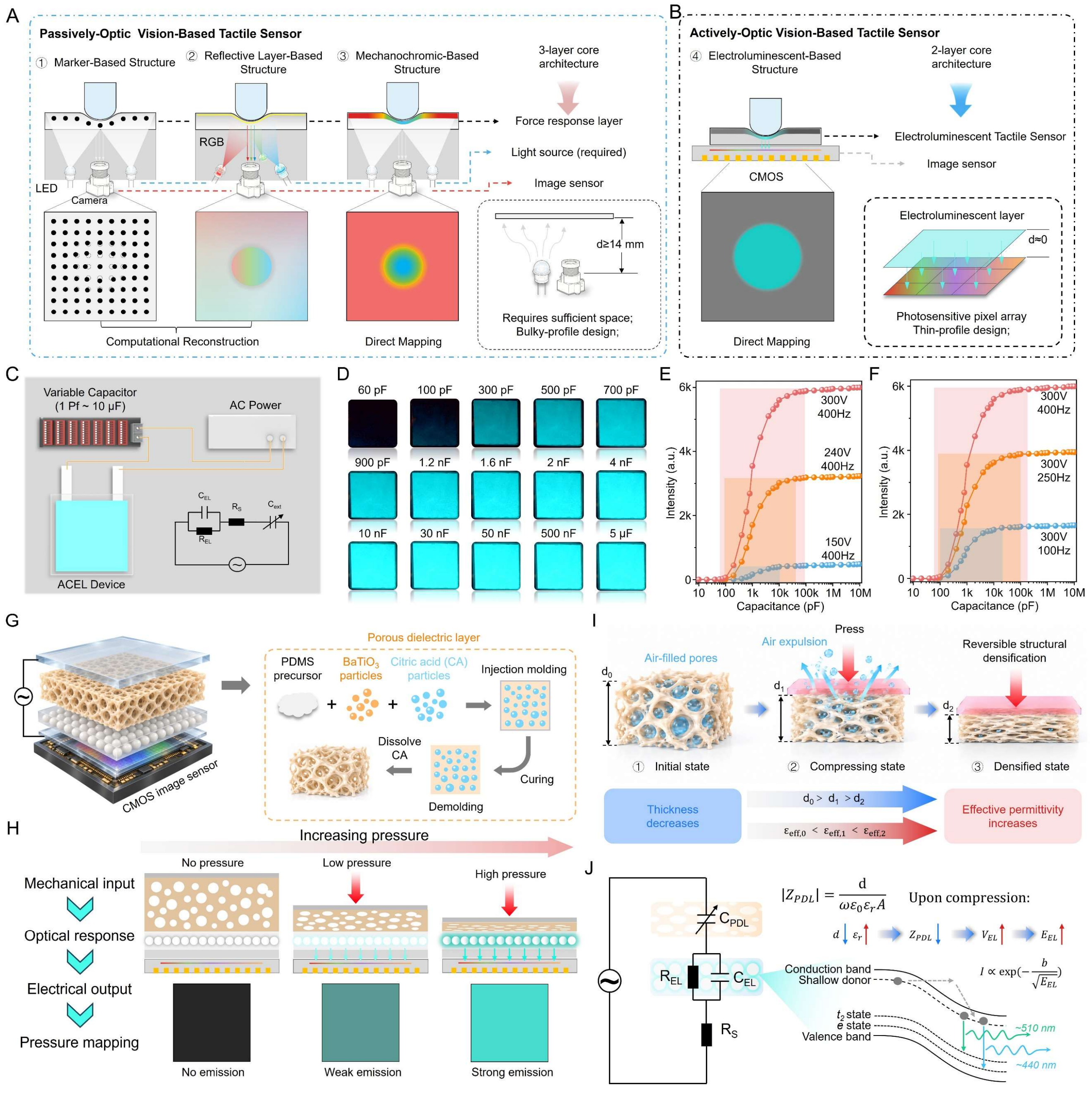

**Figure 1. Porous-dielectric ACEL for tactile imaging. (A)** Conventional vision-based tactile sensing architectures: marker-based, reflective-layer-based and mechanochromic sensing. **(B)** Proposed self-emissive tactile imaging architecture with CMOS readout. **(C)** Experimental setup and equivalent circuit for series-capacitance modulation. **(D)** ACEL emission images with different series capacitances. Capacitance-dependent emission intensity and effective modulation windows at different driving voltages **(E)** and frequencies **(F)**. **(G)** Porous-dielectric fabrication and ACEL–CMOS device architecture. **(H)** Pressure-to-light conversion and tactile imaging mechanism. **(I)** Pressure-induced pore compression, framework densification and dielectric modulation. **(J)** Equivalent circuit of the pressure-responsive ACEL device and ZnS electroluminescence mechanism.

At 300 V and 400 Hz, increasing the series capacitance progressively transformed the device from nearly non-emissive to brightly emissive (Figure 1D). The emission intensity exhibited a pronounced nonlinear dependence on capacitance (Figure 1E), remaining negligible below approximately 60 pF, increasing sharply at intermediate capacitances, and approaching saturation around 100 nF. We define the effective capacitance modulation window as the range over which capacitance variations produce appreciable changes in emission intensity. Under these conditions, the window extends approximately from 60 pF to 100 nF, providing a large tunable electrical range for pressure-responsive light emission.

The modulation window is also sensitive to the driving conditions. Reducing the voltage from 300 to 240 and 150 V at 400 Hz lowered the emission intensity and shifted the onset of detectable emission towards higher capacitance, while the intensity approached its respective plateau at lower capacitance (Figure 1E). Similarly, decreasing the frequency from 400 to 250 and 100 Hz at 300 V suppressed emission and narrowed the effective modulation range (Figure 1F). These trends arise from the combined effects of capacitive voltage division, the threshold electric field for electroluminescence, and frequency-dependent excitation. The lower boundary is primarily governed by the capacitance required to reach detectable emission, whereas the upper boundary reflects the onset of optical saturation rather than necessarily complete electrical voltage saturation.

These findings establish capacitance as a controllable parameter for regulating ACEL emission and identify the operating range required for effective pressure-to-light conversion. Replacing the external capacitor with a pressure-responsive dielectric layer could translate pressure-induced capacitance changes into local variations in emission intensity, providing the basis for spatially resolved tactile imaging.

### 2.2 Pressure-to-light transduction through compressible porous dielectrics

To translate the series-capacitance modulation mechanism into pressure-responsive electroluminescence, we incorporated a compressible porous dielectric layer into the ACEL device as an internal variable capacitor (Fig. 1G). The device comprises a transparent electrode, a ZnS-based emissive layer, a porous dielectric layer and a stretchable conductive electrode, with optical readout provided by an underlying complementary metal–oxide–semiconductor (CMOS) image sensor. The porous dielectric was fabricated from polydimethylsiloxane (PDMS), $BaTiO_3$ particles and citric acid (CA) sacrificial templates, which generate compressible pores and enable dielectric-property tuning.

The device directly encodes contact pressure into spatial variations in emission intensity (Fig. 1H). Local compression enhances electroluminescence within the contact region, producing pressure-dependent light patterns that are captured by the CMOS image sensor through the transparent electrode. This self-emissive architecture enables continuous tactile imaging without external illumination or individually addressed electrical sensing elements.

The pressure response originates from the coupled changes in dielectric geometry and permittivity (Fig. 1I). Compression reduces pore volume and dielectric thickness while increasing the effective relative permittivity of the porous composite. Consequently, the capacitance increases according to:

$$C_{PDL}(P) = \frac{\varepsilon_0 \varepsilon_{eff}(P) A}{d(P)} \tag{4}$$

where $\varepsilon_0$ is the vacuum permittivity, $A$ is the effective electrode overlap area, and $\varepsilon_{eff}(P)$ and $d(P)$ represent the pressure-dependent effective relative permittivity and dielectric thickness, respectively.

The resulting capacitance increase reduces the dielectric reactance and redistributes the applied AC voltage towards the emissive layer, as described by the equivalent circuit (Fig. 1J):

$$V_{EL}(P) = V_{AC} \frac{Z_{EL}}{R_S + Z_{EL} + \frac{1}{j\omega C_{PDL}(P)}} \tag{5}$$

Within the operating range, the increased voltage across the ZnS-based emissive layer enhances its local electric field and electroluminescence. Thus, pressure-induced dielectric modulation directly converts local mechanical compression into spatially resolved light emission, providing a continuous pressure-imaging mechanism without discrete sensing pixels.

### 2.3 Effects of dielectric composition and structure on electroluminescence

The porous dielectric layer was first examined to establish its structural and compositional characteristics (**Fig. 2A**). Scanning electron microscopy (SEM) revealed interconnected pores formed by removing the citric acid sacrificial template, while energy-dispersive X-ray spectroscopy (EDS) mapping confirmed the incorporation of $BaTiO_3$ fillers into the PDMS matrix. To assess dielectric tunability, we measured the pressure-dependent capacitance of porous layers with different $BaTiO_3$ loadings (Fig. 2B). All samples exhibited nonlinear capacitance increases under compression, with higher filler contents producing larger initial capacitances and pressure-induced changes. At a $BaTiO_3$-to-PDMS ratio of 2:1, capacitance increased from approximately 20 to 235 pF ($\Delta C$ = 215 pF), compared with 11 to 56 pF ($\Delta C$ = 45 pF) without $BaTiO_3$ under the same pressure range. This 4.8-fold enhancement in capacitance modulation provides a means of tuning the dielectric response towards the effective ACEL modulation window.

The corresponding pressure-to-light conversion was investigated using ACEL devices incorporating different porous dielectrics, with optical readout by a CMOS image sensor (Fig. 2C). Emission intensity increased with pressure across the investigated range of 0-1.70 MPa, with pronounced differences among compositions (Fig. 2D). The $BaTiO_3$-free device remained weakly emissive even at 1.70 MPa, whereas the $BaTiO_3$-to-PDMS ratio of 2:1 composition produced detectable emission at 56 kPa and progressively brighter emission under increasing pressure. Spectroscopic measurements of this composition revealed increasing emission intensity with compressive strain, while the characteristic emission peak remained near 502 nm (Fig. 2E). Pseudocolor imaging further visualized the pressure-dependent emission patterns (bottom panel of Fig. 2D). Grayscale signals extracted from CMOS images followed the same trend as spectroscopically measured emission intensities (Fig. 2F), enabling quantitative optical readout by CMOS image sensors. The 2:1 composition exhibited the strongest pressure-dependent grayscale response among the tested samples, consistent with its larger capacitive modulation (Fig. 2B).

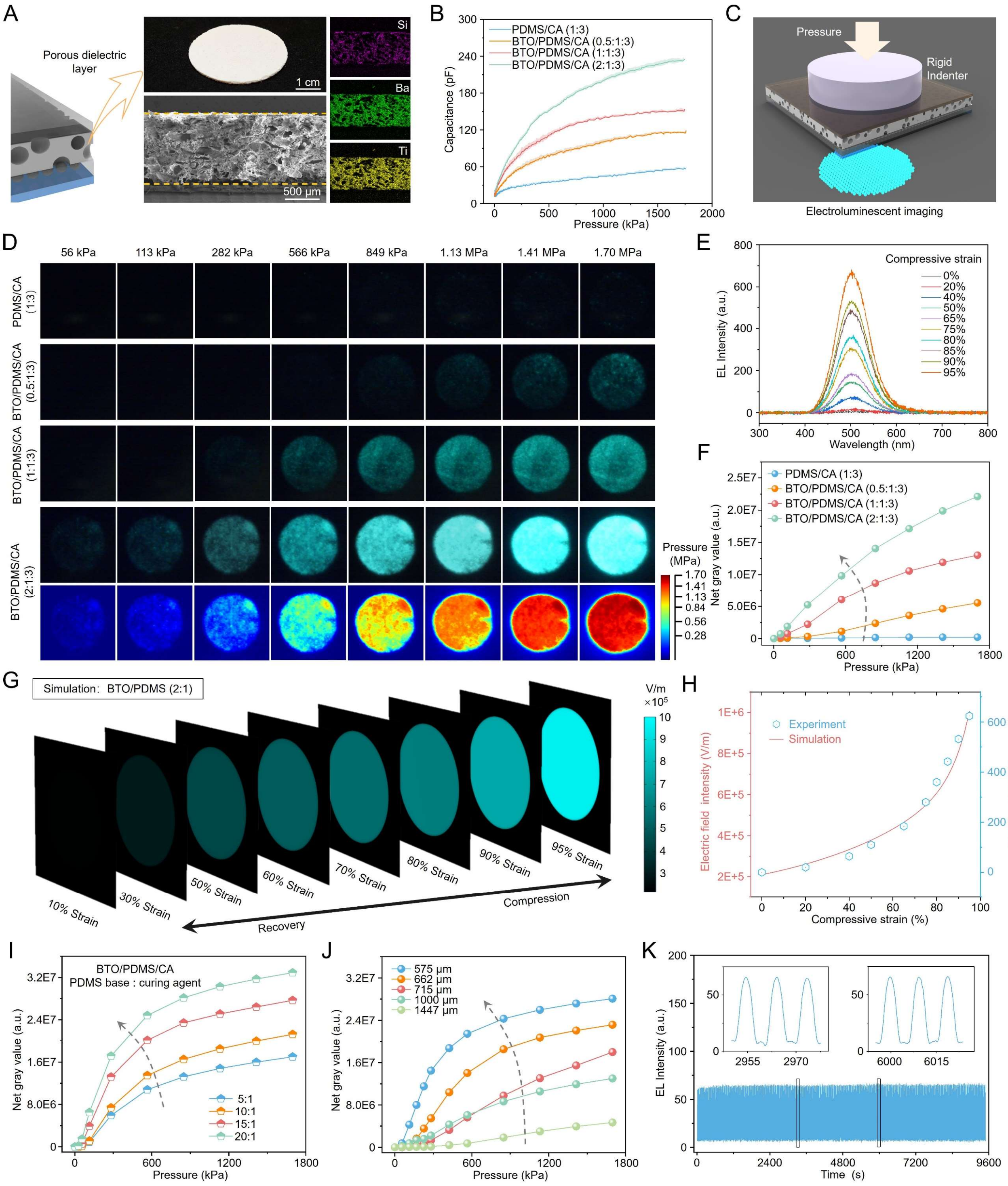

**Figure 2. Effects of composition and thickness on electroluminescence.** (**A**) Photograph, cross-sectional SEM image and EDS elemental maps of the porous dielectric. (**B**) Pressure-dependent capacitance at different $BaTiO_3$ loadings. (**C**) Experimental setup for pressure-modulated electroluminescence and CMOS imaging. (**D**) Electroluminescence images and pseudocolor intensity maps at different pressures and $BaTiO_3$ loadings. (**E**) Emission spectra under increasing compressive

strain. (**F**) Pressure-dependent net grayscale responses at different $BaTiO_3$ loadings. (**G**) Simulated electric-field distributions in the emissive layer under compression. (**H**) Simulated electric-field intensity and measured emission intensity versus compressive strain. Effect of the PDMS base-to-curing-agent ratio (**I**) and dielectric thickness (**J**) on pressure-responsive emission. (**K**) Pressure sensing stability under cyclic loading.

To examine the underlying mechanism, finite-element simulations were performed using experimentally derived capacitance and deformation parameters (Fig. 2G). Compression increases the effective permittivity and reduces the thickness of the porous dielectric, lowering its capacitive reactance and redistributing the applied voltage towards the emissive layer. The simulated electric-field enhancement followed the experimentally observed increase in emission intensity (Fig. 2H), supporting dielectric impedance modulation as the dominant mechanism of pressure-responsive electroluminescence.

The optical response could be further tuned through the mechanical compliance and thickness of the porous dielectric. Increasing the PDMS base-to-curing-agent ratio from 5:1 to 20:1 enhanced the net grayscale response by ~ 2 times at 1.70 MPa (Fig. 2I), consistent with greater compression of the more compliant dielectric layer. Reducing the dielectric thickness from 1447 to 575 μm increased the corresponding response by ~ 6 times (Fig. 2J), consistent with increased capacitance and reduced dielectric reactance. The device also maintained reproducible periodic emission over approximately 9600 s of cyclic operation without noticeable signal degradation, indicating the high robustness of the sensing device (Fig. 2K).

Together, these results establish dielectric composition, mechanical compliance and layer thickness as key parameters for regulating pressure-to-light conversion. Their combined control provides considerable flexibility in tuning the pressure-responsive electroluminescence for high-resolution tactile imaging.

### 2.4 Pore-size engineering for uniform tactile imaging

Although porous dielectrics enable pressure-responsive electroluminescence, their heterogeneous microstructures can compromise spatial emission uniformity. To investigate this effect, we developed three-dimensional finite-element models incorporating the solid framework and air-filled pores with characteristic sizes of approximately 600 and 100 μm (**Figs**. **3A**, 3B). In the large-pore structure, pronounced electric-field heterogeneity persisted even at 95% compressive strain, with localized low-field regions remaining within the emissive layer (Fig. 3A). By contrast, reducing the pore size to approximately 100 μm produced a more continuous electric-field distribution, progressively suppressing low-field regions as compressive strain increased to 90% (Fig. 3B). These results suggest that pore size governs not only the pressure-induced electric-field enhancement but also its spatial uniformity.

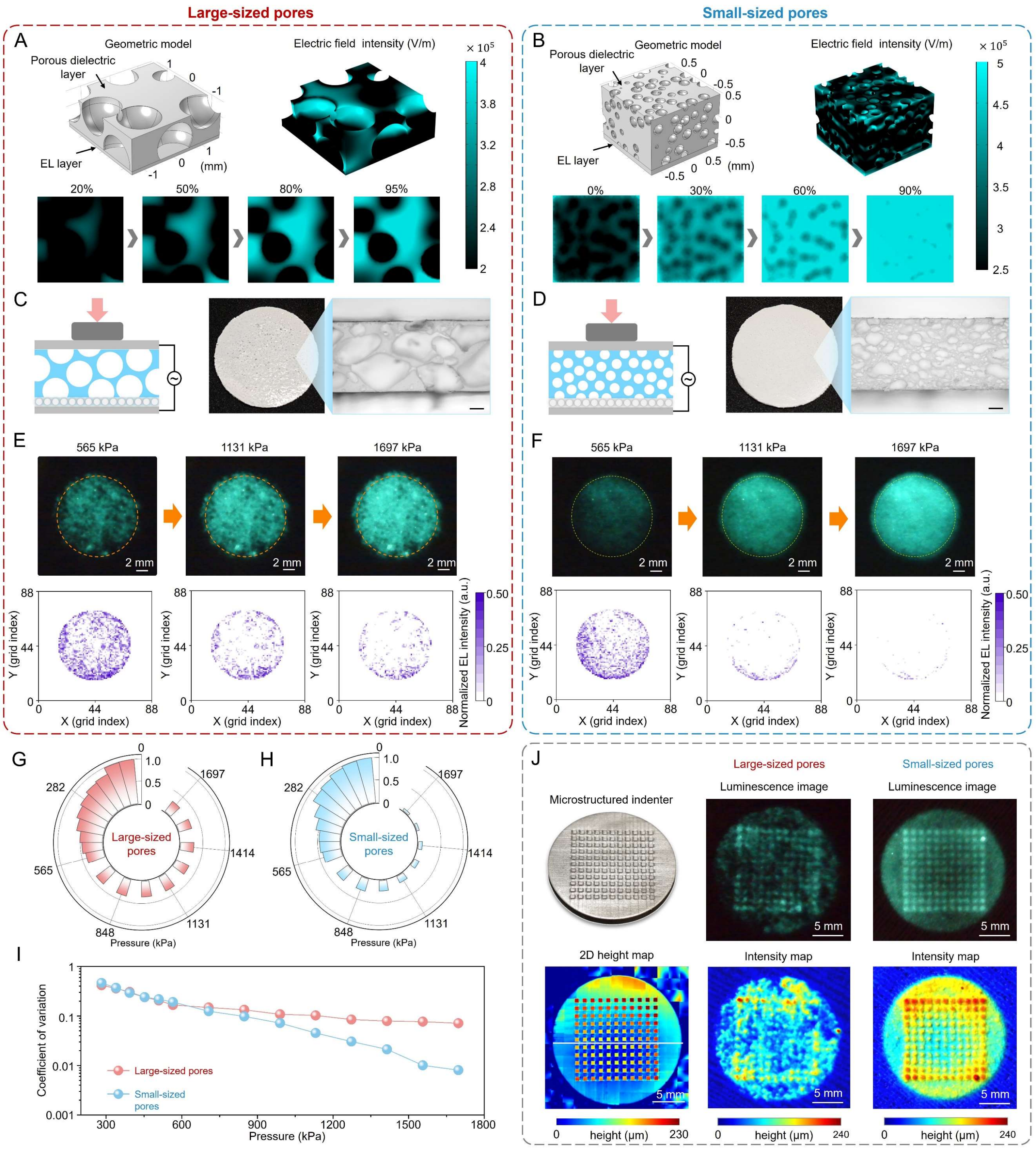

**Figure 3. Pore-size-dependent electroluminescence.** Simulated electric-field distributions under compression for large-pore (**A**) and small-pore (**B**) dielectrics. Structural schematics, photographs and cross-sectional SEM images of large-pore (**C**) and small-pore (**D**) dielectrics. Pressure-dependent electroluminescence images and extracted non-emissive regions for large-pore (**E**) and small-pore (**F**) devices. Non-emissive area fractions versus applied pressure for large-pore (**G**) and small-pore (**H**) devices. (**I**) Pressure-dependent coefficient of variation (CV) of spatial emission intensity for both devices. (**J**) Surface topography of the square-array indenter and corresponding tactile images and pseudocolor intensity maps obtained using devices with different pore sizes.

To verify this prediction, porous dielectrics with distinct pore sizes were fabricated and characterized by optical imaging and cross-sectional SEM (Figs. 3C, 3D). Their pressure-dependent electrolumine-scent patterns were then compared at 565, 1131 and 1697 kPa (Figs. 3E, 3F). Although the large-pore device became brighter under increasing pressure, numerous non-emissive regions persisted even at 1697 kPa (Fig. 3E). In contrast, the small-pore device exhibited increasingly continuous emission, with substantially fewer dark regions and improved spatial uniformity (Fig. 3F). Image segmentation of the non-emissive regions (lower panels of Figs. 3E, 3F) further confirmed the distinct spatial emission patterns, consistent with the simulated electric-field distributions.

We quantitatively analyzed the non-emissive area fraction under different pressures (Figs. 3G, 3H). Both devices were non-emissive without applied pressure, whereas compression progressively expanded the emissive regions. For the large-pore device, the reduction in non-emissive area slowed markedly above approximately 900 kPa, leaving persistent emission deficiencies at higher pressures. In contrast, the small-pore device showed a continuous reduction in non-emissive area, consistent with the more uniform electric-field distributions predicted by the simulations. The coefficient of variation (CV) further quantified emission uniformity (Fig. 3I). At approximately 1.70 MPa, the CV decreased from 0.08 for the large-pore device to 0.009 for the small-pore device, representing an approximately ninefold improvement in spatial intensity uniformity.

We further evaluated the impact of emission uniformity on tactile imaging fidelity by reconstructing a laser-patterned square array with 600 μm × 600 μm features (Fig. 3J). The large-pore device exhibited missing features and distorted patterns caused by localized emission deficiencies. In contrast, the small-pore device reproduced the array with greater continuity and closer spatial correspondence to the indenter topography. These results establish pore-size engineering as an effective means of improving both emission uniformity and pressure-pattern fidelity, linking dielectric microstructure to the quality of optical tactile imaging.

### 2.5 High-resolution vision-based tactile sensing

Spatially resolved tactile imaging requires distinct optical responses from adjacent contact regions rather than merely enhanced overall emission. We investigated this capability using microstructured indenters pressed against the continuous ACEL interface (**Fig. 4A**). Protruding features locally compress the porous dielectric, whereas adjacent recessed regions undergo less deformation. Finite-element simulations reveal that this differential compression generates spatial variations in electric potential and enhances the electric field preferentially beneath the contact regions (Fig. 4B). The resulting contrast is further amplified by the nonlinear dependence of electroluminescence on the local electric field (Fig. 4C). Thus, spatially selective dielectric modulation enables continuous ACEL interfaces to encode fine contact features into distinct emission-intensity patterns without individually addressed sensing pixels.

We evaluated the spatial resolving capability using striped indenters with feature widths of 100, 200 and 500 μm (Figs. 4D-4F). For 100 μm stripes, increasing the applied load from 10 to 50 N enhanced emission contrast and progressively revealed distinct periodic patterns corresponding to the indenter topography (Fig. 4D). Features measuring 200 and 500 μm were likewise faithfully reproduced, with clearly distinguishable emission boundaries (Figs. 4E, 4F). These results demonstrate direct optical imaging of contact features down to the 100 μm scale through localized electroluminescence modulation.

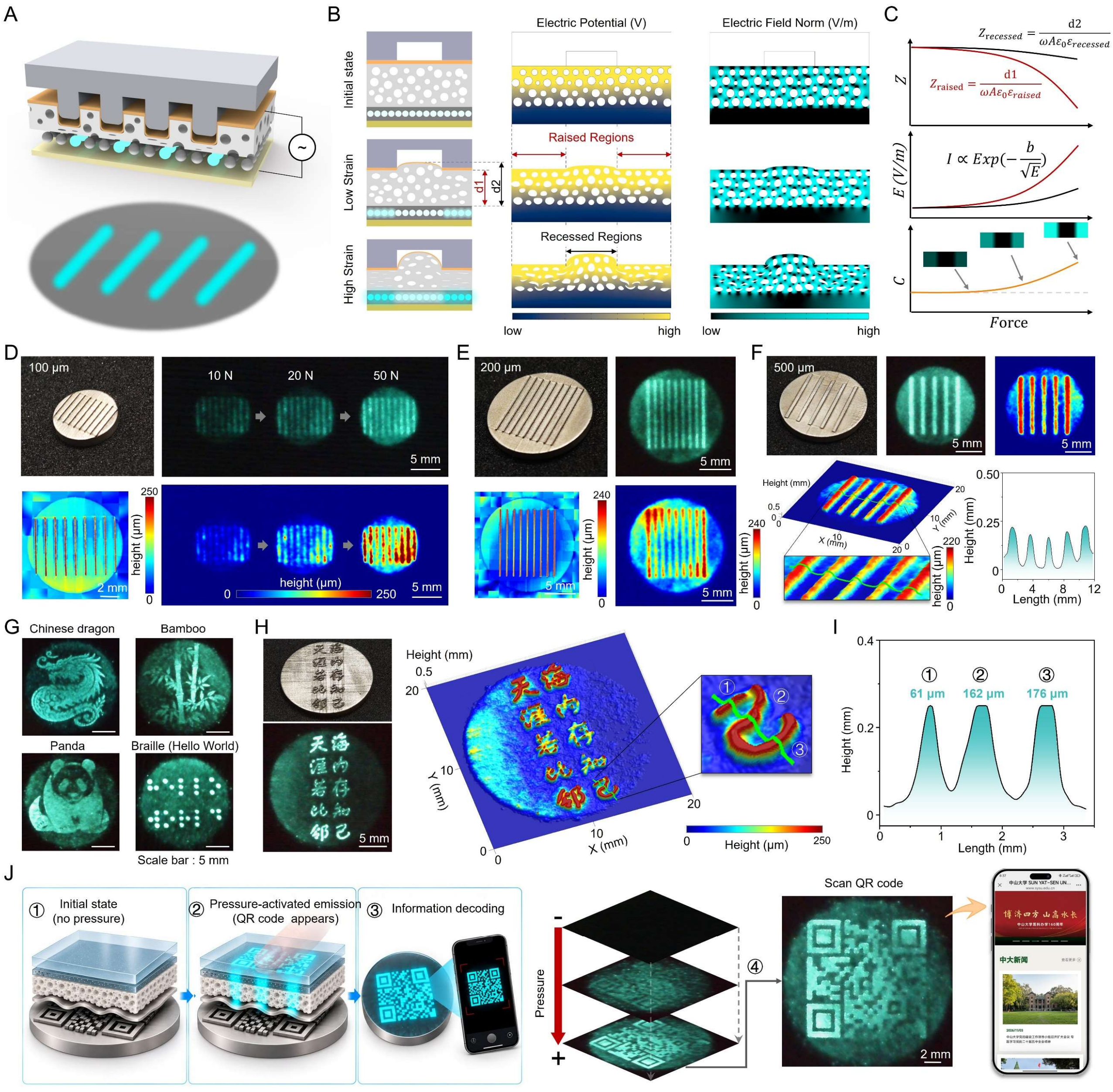

**Figure 4. High-resolution tactile imaging**. (A) Schematic of spatially selective electroluminescence under microstructured contact. (B) Simulated deformation, electric potential and electric-field distributions under compression. (C) Local dielectric impedance modulation and nonlinear electroluminescent response. (D) Electroluminescence images and pseudocolor intensity maps of 100 μm stripes under different loads. (E) Surface topography and tactile images of 200 μm stripes. (F) Surface topography, tactile images and height profiles of 500 μm stripes. (G) Tactile images of Chinese dragon, bamboo, panda and Braille patterns. (H) Surface topography of the Chinese-character mold and corresponding tactile images. (I) Cross-sectional height profile of adjacent microstructures. (J) Pressure-triggered QR-code display and optical readout.

The device also reproduced complex contact patterns, including Chinese dragon, bamboo, panda and Braille structures, preserving curved contours, fine features and spatially separated raised dots (Fig. 4G). More demanding patterns comprising densely arranged Chinese characters ("海内存知己，天涯若比邻") were resolved with recognizable strokes and structural details (Fig. 4H). Intensity profiles across representative features revealed three distinct peaks separated by low-intensity regions (Fig. 4I), confirming the spatial distinguishability of adjacent contact structures. These demonstrations establish the ability of the continuous emissive interface to preserve complex spatial contact information beyond regular microstructured patterns.

Beyond tactile imaging, we demonstrated pressure-gated optical information readout using a QR-code pattern (Fig. 4J). Mechanical compression activated the electroluminescent pattern, revealing a previously weakly visible code that could be recognized by a smartphone. This demonstration extends pressure-responsive electroluminescence to mechanically controlled optical information display and readout. Together, these results establish spatially selective dielectric modulation as a route to high-resolution tactile imaging and interactive optical interfaces.

## 3. Conclusion

This work establishes dielectric-layer porosification in ACEL devices as a strategy for directly encoding contact pressure distributions into spatially resolved light emission. While dielectric composition, thickness and mechanical properties regulate the pressure-dependent electroluminescent response, pore architecture further influences spatial emission uniformity and imaging fidelity. By optimizing these parameters, the continuous ACEL interface resolves contact features down to 100 μm without discrete sensing pixels or external illumination.

Although direct CMOS readout has been demonstrated, the imaging performance of fully integrated ACEL-CMOS devices remains to be systematically evaluated, particularly in terms of spatial resolution, mechanical robustness and operational stability. Future advances in ultrathin, flexible and large-area image sensors could further reduce device thickness and extend the sensing area while preserving the mechanical compliance and scalability of ACEL interfaces. Such developments may ultimately shift vision-based tactile sensing from bulky, geometry-constrained optical assemblies towards thin, conformable and large-area sensing interfaces, offering greater freedom in the design of dexterous robotic hands.

## 4. Experimental Section

**Preparation of Porous Dielectric Layers**: $BaTiO_3$/PDMS composite porous dielectric layers were fabricated using a sacrificial templating method. Polydimethylsiloxane (PDMS, Sylgard 184) served as the flexible polymer matrix, $BaTiO_3$ particles as high-permittivity fillers, and citric acid particles as water-soluble sacrificial templates. Initially, $BaTiO_3$ particles (3.0 g), PDMS base (1.5 g), and citric acid particles (4.5 g) were weighed at a mass ratio of 2:1:3. The $BaTiO_3$ particles were gradually added to the PDMS base under mechanical stirring to ensure uniform dispersion, yielding a homogeneous $BaTiO_3$/PDMS composite mixture. Subsequently, 0.15 g of Sylgard 184 curing agent was introduced at a PDMS base-to-curing agent mass ratio of 10:1, followed by further stirring until thoroughly mixed. The preweighed citric acid particles were then gradually incorporated under continuous mechanical stirring to obtain a homogeneous composite precursor containing dispersed sacrificial templates. The resulting mixture was sandwiched between two glass plates ($5 \times 5$ cm$^2$), with the plate spacing adjusted to control the initial film thickness to 1 mm. The assembly was secured with clamps to maintain a consistent thickness during curing and subsequently heated in an oven at 80 °C for 2 h, allowing the PDMS matrix to crosslink and form a composite elastomeric film containing embedded $BaTiO_3$ and citric acid particles. After curing, the composite film was carefully separated from the glass plates and immersed in a mixture of absolute ethanol and deionized water. Repeated ultrasonic washing with periodic replacement of the washing solution was performed to promote the dissolution and removal of the citric acid templates. As the sacrificial particles dissolved, interconnected or isolated pores were formed within the PDMS matrix, ultimately yielding a compressible $BaTiO_3$/PDMS porous dielectric layer for subsequent ACEL device assembly and characterization.

**Preparation of Electroluminescent Layers**: ZnS/PDMS composite electrolumine-scent layers were fabricated using a doctor-blading method, with copper-doped zinc sulfide (ZnS) particles serving as the electroluminescent phosphor and polydimethyl-siloxane (PDMS, Sylgard 184) as the flexible polymer matrix. Initially, 5.0 g of PDMS base and 15.0 g of ZnS phosphor powder were weighed at a mass ratio of 1:3. The ZnS powder was gradually added to the PDMS base and mechanically stirred at 500 rpm for 10 min to ensure uniform dispersion of the phosphor particles within the polymer matrix, yielding a homogeneous composite mixture. Subsequently, 0.5 g of Sylgard 184 curing agent was added at a PDMS base-to-curing agent mass ratio of 10:1, followed by further stirring at 500 rpm for 5 min to obtain a uniformly mixed, thermally curable ZnS/PDMS slurry. The resulting slurry was uniformly deposited onto a precleaned glass substrate using a doctor blade, with the blade gap adjusted to control the wet-film thickness to 200 μm. The coated sample was subsequently cured in an oven at 80 °C for 2 h, allowing the PDMS matrix to crosslink and form a flexible composite electroluminescent layer with ZnS phosphor particles dispersed throughout the polymer network. The resulting layer was used for subsequent ACEL device assembly.

**Fabrication of Stretchable Electrodes**: PDMS-based stretchable electrodes were fabricated by combining spin coating and screen printing. First, PDMS base (Sylgard 184) and its curing agent were mixed at a mass ratio of 10:1, followed by thorough mechanical stirring and vacuum degassing to remove entrapped air bubbles. The degassed PDMS precursor was then deposited onto a cleaned glass substrate and spin-coated at high speed to form a uniform thin film. The spin-coating parameters were adjusted to achieve a film thickness of approximately 50 μm, followed by thermal curing to obtain an ultrathin, flexible PDMS substrate. To improve the interfacial adhesion between the conductive silver paste and the PDMS substrate, the cured PDMS surface was subjected to surface activation. The glass-supported PDMS film was subsequently mounted on a screen-printing platform, and a stretchable conductive silver paste was uniformly printed onto the PDMS surface using a squeegee. The printing parameters were adjusted to control the thickness and uniformity of the conductive layer, forming a continuous, planar conductive network. Following printing, the samples were thermally treated according to the curing requirements of the silver paste to establish stable electrical conduction pathways. After cooling, the composite film was carefully peeled off from the glass substrate, yielding a stretchable electrode consisting of an ultrathin PDMS elastomer substrate and a screen-printed conductive silver layer.

**Assembly of ACEL Tactile Sensors**: Multilayer ACEL tactile sensors were fabricated through layer-by-layer assembly and flexible encapsulation. First, a transparent indium tin oxide-coated polyethylene terephthalate (ITO/PET) substrate with a preformed ZnS/PDMS composite electroluminescent layer was used as the bottom functional unit. The $BaTiO_3$/PDMS porous dielectric layer, after removal of the sacrificial templates and thorough drying, was cut to the desired dimensions and carefully laminated onto the electroluminescent layer. A low preloading pressure was applied during lamination to minimize initial compression of the porous structure while avoiding noticeable interfacial air gaps. Subsequently, the prefabricated PDMS/stretchable silver composite electrode was placed on top of the porous dielectric layer, with the screen-printed conductive silver surface facing the dielectric layer and the PDMS elastomer substrate facing outward. This assembly formed a multilayer device comprising, from bottom to top, an ITO/PET bottom electrode, a ZnS/PDMS electroluminescent layer, a $BaTiO_3$/PDMS porous dielectric layer, and a stretchable silver top electrode. After alignment and lamination, flexible silicone adhesive was applied around the perimeter of the active sensing area for interlayer fixation and edge encapsulation, while preventing the encapsulation material from penetrating the porous dielectric structure. Electrical leads were then connected to the bottom ITO/PET electrode and the top stretchable silver electrode to establish the AC driving circuit. Finally, the assembled devices were inspected for interfacial contact, electrical insulation, and initial electrical response, and their pressure-dependent electroluminescent behavior was verified under AC excitation before subsequent vision-based tactile imaging experiments.

**Pressure-Dependent Electroluminescence Characterization**: The pressure-dependent electro-luminescent performance of the ACEL devices was characterized using a universal testing machine (TSMT EM2.502) equipped with a force sensor (Transcell BSA-50kg). During testing, the device was horizontally positioned at the center of the testing platform, with a CMOS image sensor (HBV-1917HD V33, 1920 × 1080 pixels, 30 fps) integrated underneath to capture the electroluminescent images. Normal loads were applied using a cylindrical stainless-steel flat-ended indenter with a diameter of 15 mm. The loading rate was set to 5 mm $min^{-1}$, with a maximum applied force of 300 N, corresponding to a nominal contact pressure of approximately 1.70 MPa. During loading, the applied force was continuously monitored by the force sensor, while electroluminescent images were synchronously acquired and recorded at 10 fps. To quantitatively characterize the pressure-dependent optical response, electroluminescence spectra were recorded using a fiber-optic spectrometer (Ocean Optics USB2000, 340-1020 nm). The optical fiber probe was maintained at a fixed position relative to the emitting region, with a collection distance of approximately 5 mm. The spectrometer was operated with an integration time of 100 ms, three spectral averages, and a smoothing setting of 5. Spectra were collected at 15 selected pressure levels over the range of 0-1.70 MPa. At each pressure level, the applied load was maintained for 3 s before spectral acquisition to allow the force reading to stabilize. Measurements were repeated three times at each pressure level, and the resulting spectra were averaged. The spectral data were exported for subsequent analysis of emission peak position, peak intensity, and integrated emission intensity as functions of applied pressure.

**Patterned Tactile Imaging and Spatial Resolution Characterization**: Patterned stainless-steel molds containing fine microstructures were employed to evaluate the spatial resolution and pattern imaging capability of the ACEL-CMOS vision-based tactile sensing system. During testing, the patterned molds were positioned between the indenter of the universal testing machine and the ACEL device, with the microstructured surfaces facing and contacting the top stretchable electrode. Normal loads were applied through the testing machine, generating spatially varying local pressure distributions corresponding to the microstructural features of the molds and thereby inducing spatially resolved electroluminescent responses. The resulting emission patterns were captured in real time through the transparent ITO/PET electrode using the CMOS image sensor (HBV-1917HD V33, 1920 × 1080 pixels, 30 fps) integrated beneath the device. The acquired images were subsequently processed through grayscale conversion and pseudocolor mapping to visualize the spatial distribution of electroluminescent intensity. By comparing the reconstructed optical patterns with the actual surface topography of the molds, the imaging performance was evaluated in terms of pattern fidelity, distinguishability of adjacent features, and spatial correspondence, providing an assessment of the system's capability to resolve fine contact features.

**Finite Element Simulations**: Finite element simulations were performed using COMSOL Multiphysics to investigate the mechanical deformation, electric-field evolution, and spatial electroluminescence modulation of porous dielectric structures under external loading. Geometrical models were constructed according to the device architecture, and the corresponding mechanical and dielectric properties, including Young's modulus, Poisson's ratio, and relative permittivity, were assigned to the constituent materials. In the Solid Mechanics module, the bottom boundary was fixed, while a normal load or prescribed displacement was applied to designated regions of the top surface to simulate structural deformation under different loading conditions. For electric-field analysis, the bottom electrode was grounded, and a prescribed electric potential was applied to the top electrode to calculate the electric potential and field distributions at different deformation states. To investigate the pressure-induced enhancement of the overall electric field, an effective-medium model was employed to represent the porous dielectric layer. Its effective dielectric properties at different compression states were determined from experimentally measured pressure-dependent capacitance and thickness variations, allowing the influence of dielectric compression on the electric field within the electroluminescent layer to be evaluated. To elucidate the influence of pore size on spatial electric-field uniformity, a three-dimensional geometrical model explicitly incorporating the PDMS composite skeleton and air-filled pores was further constructed. By varying the pore size, the local electric-field distributions and the evolution of low-field regions were compared under different compression states. Additionally, a localized contact model comprising a microstructured indenter, a porous dielectric layer, and an electroluminescent layer was established to investigate deformation, electric potential distribution, and electric-field responses induced by patterned mechanical loading. Particular attention was paid to the spatial electric-field contrast between contacting and adjacent non-contacting regions. Collectively, these simulations provided insights into three aspects of the electromechanical response: pressure-induced electric-field enhancement, pore-size-dependent field uniformity, and localized electric-field modulation, thereby supporting the interpretation of pressure-to-light transduction and high-resolution tactile imaging in ACEL devices.

**Completing interest:** All authors declare no completing interest.

**Acknowledgments:** This work is supported by National Natural Science Foundation of China (22475242), the Guangdong Basic and Applied Basic Research Foundation (2025A1515010271), and the GBRCE for Functional Molecular Engineering (31000-42080003).

**Author contributions:** J.G. conceived and supervised the project. M.W.G. led the overall research and was responsible for experimental design, material preparation, device fabrication, performance characterization, finite element simulations, system integration, high-resolution tactile imaging, data analysis, and interpretation. L.C.L. and Z.Y.Z. contributed to material characterization and device performance testing. Y.Y. and Z.X.N. assisted with the development and assembly of experimental setups. X.Y.M. and J.S.C. contributed to the conceptualization and design of the main figures. L.Z. provided technical support for high-resolution tactile imaging, while J.H.G. assisted with scanning electron microscopy characterization. J.S.X. contributed to device fabrication and experimental assistance. Z.L.W. and M.X.H. provided additional assistance with experimental work and data processing. M.W.G. wrote the original draft of the manuscript, and J.G. revised the manuscript. All authors discussed the results, reviewed the manuscript, and approved the final version.

**Data availability:** The data that support the findings of this study are available from the corresponding author upon reasonable request.

**Reference**

[1] Z. Zhao, W. Li, Y. Li, T. Liu, B. Li, M. Wang, K. Du, H. Liu, Y. Zhu, Q. Wang, K. Althoefer, S.-C. Zhu, *Nature Machine Intelligence* **2025**, 7, 889.

[2] Q. Mao, Z. Liao, J. Yuan, R. Zhu, *Nature Communications* **2024**, 15, 6871.

[3] Y. Yan, A. Zermane, J. Pan, A. Kheddar, *Nature Machine Intelligence* **2024**, 6, 1284.

[4] C. Son, J. Kim, D. Kang, S. Park, C. Ryu, D. Baek, G. Jeong, S. Jeong, S. Ahn, C. Lim, Y. Jeong, J. Eom, J.-H. Park, D. W. Lee, D. Kim, J. Kim, H. Ko, J. Lee, *Nature Communications* **2024**, 15, 8003.

[5] N. Bai, Y. Xue, S. Chen, L. Shi, J. Shi, Y. Zhang, X. Hou, Y. Cheng, K. Huang, W. Wang, J. Zhang, Y. Liu, C. F. Guo, *Nature Communications* **2023**, 14, 7121.

[6] Q. Nie, F. Wang, F.-S. Yang, H. Xun, J. Hou, Q. Xu, Y. Hong, J. Zhang, X. Wei, Y.-F. Lin, P.-W. Chiu, L. Zeng, M. Li, B. Wang, J. Zhang, *Advanced Materials* **2026**, 38, e10646.

[7] Y. Chen, J. Cao, J. Qiu, D. Yang, M. Liu, M. Zhang, C. Li, Z. Wu, J. Yu, X. Zhang, X. Chen, Z. Huang, E. Song, M. Wang, Q. Liu, M. Liu, *Nature Communications* **2025**, 16, 5691.

[8] B. Shao, M.-H. Lu, T.-C. Wu, W.-C. Peng, T.-Y. Ko, Y.-C. Hsiao, J.-Y. Chen, B. Sun, R. Liu, Y.-C. Lai, *Nature Communications* **2024**, 15, 1238.

[9] X. Tian, G. Cheng, Z. Wu, X. Wen, Y. Kong, P. Long, F. Zhao, Z. Li, D. Zhang, Y. Hu, D. Wei, *Advanced Functional Materials* **2024**, 34, 2406022.

[10] D. Kong, Y. Lu, S. Zhou, M. Wang, G. Pang, B. Wang, L. Chen, X. Huang, H. Lyu, K. Xu, G. Yang, *Science Advances*, 11, eadv2124.

[11] B. Lee, J.-Y. Oh, H. Cho, C. W. Joo, H. Yoon, S. Jeong, E. Oh, J. Byun, H. Kim, S. Lee, J. Seo, C. W. Park, S. Choi, N.-M. Park, S.-Y. Kang, C.-S. Hwang, S.-D. Ahn, J.-I. Lee, Y. Hong, *Nature Communications* **2020**, 11, 663.

[12] Z. Wu, S. Chen, S. Fan, Z. Qiao, J. Qi, Y. Zhang, C. T. Lim, *Nature Communications* **2026**, 17, 6886.
[13] K. Qiao, X. Feng, L. Dong, L. Liu, Q. Ma, D. Zhao, S. Lu, H. Wang, C. Yu, L. Pan, B. Wang, *Nature Communications* **2026**, 17, 3928.
[14] D. Zhong, C. Wu, Y. Jiang, Y. Yuan, M.-g. Kim, Y. Nishio, C.-C. Shih, W. Wang, J.-C. Lai, X. Ji, T. Z. Gao, Y.-X. Wang, C. Xu, Y. Zheng, Z. Yu, H. Gong, N. Matsuhisa, C. Zhao, Y. Lei, D. Liu, S. Zhang, Y. Ochiai, S. Liu, S. Wei, J. B. H. Tok, Z. Bao, *Nature* **2024**, 627, 313.
[15] K. Park, H. Yuk, M. Yang, J. Cho, H. Lee, J. Kim, *Science Robotics*, 7, eabm7187.
[16] K. Kim, J.-H. Hong, K. Bae, K. Lee, D. J. Lee, J. Park, H. Zhang, M. Sang, J. E. Ju, Y. U. Cho, K. Kang, W. Park, S. Jung, J. W. Lee, B. Xu, J. Kim, K. J. Yu, *Science Advances*, 10, eadr1099.
[17] Z. Li, B. Da, J. Shi, T. Wang, W. Si, B. Li, H. He, Z. Liao, T. Wu, X. Han, L. Bai, H. Kong, C. Song, C. F. Guo, *Nature Machine Intelligence* **2026**, DOI: 10.1038/s42256-026-01325-6.
[18] Y. Li, J. Long, Y. Chen, Y. Huang, N. Zhao, *Advanced Materials* **2022**, 34, 2200517.
[19] H. Sun, A. Spiers, H. Lee, J. Fiene, G. Martius, *Nature Communications* **2025**, 16, 8031.
[20] A. Ortone, M. Filosa, G. Indiveri, G. Desoli, A. Mazzoni, C. M. Oddo, *Nature Communications* **2026**, 17, 2108.
[21] H. Park, K. Park, S. Mo, J. Kim, *IEEE Transactions on Robotics* **2021**, 37, 1570.
[22] H. Sun, K. J. Kuchenbecker, G. Martius, *Nature Machine Intelligence* **2022**, 4, 135.
[23] N. Zhang, J. Ren, Y. Dong, X. Yang, R. Bian, J. Li, G. Gu, X. Zhu, *Nature Communications* **2025**, 16, 2395.
[24] Z. Chen, N. Ou, X. Zhang, Z. Wu, Y. Zhao, Y. Wang, E. S. Papastavridis, N. Lepora, L. Jamone, J. Deng, S. Luo, *Nature Communications* **2026**, 17, 2101.
[25] L. Chen, C. Hegde, S. Das, W. Zhang, R. Y. Teo, J. Y. E. Chan, N. He, F. Fan, H. Duan, L. H. Wong, U. Levy, S. Magdassi, J. K. W. Yang, *Advanced Functional Materials* **2026**, 36, e24139.
[26] G. Sasso, A. Pagani, A. M. Duncan, G. Pedrizzetti, N. Pugno, J. J. C. Busfield, F. Carpi, *Science Advances* **2026**, 12, eaee5236.
[27] X. Qiu, J. Liu, B. Zhou, X. Zhang, *Advanced Functional Materials* **2023**, 33, 2300321.
[28] P. Zhang, Y. Wu, Z. Guo, Y. Song, J. Hu, H. Liu, L. Zhao, *Advanced Functional Materials* **2025**, 35, 2412911.
[29] K.-W. Sou, W.-S. Chan, K.-C. Lei, Z. Wang, S. Li, D. Peng, W. Ding, *Advanced Functional Materials* **2025**, 35, 2420872.
[30] W. Jiang, S. Lee, G. Zan, K. Zhao, C. Park, *Advanced Materials* **2024**, 36, 2304053.
[31] H. Yin, Y. Zhu, K. Youssef, Z. Yu, Q. Pei, *Advanced Materials* **2022**, 34, 2106184.
[32] Z. Liu, Z. Liu, X. Lei, Y. Yang, T. Jing, *Nature Communications* **2025**, 16, 11190.
[33] J. Ji, I. F. Perepichka, J. Bai, D. Hu, X. Xu, M. Liu, T. Wang, C. Zhao, H. Meng, W. Huang, *Nature Communications* **2021**, 12, 54.
[34] X. Fu, G. Wan, H. Guo, H.-J. Kim, Z. Yang, Y. J. Tan, J. S. Ho, B. C. K. Tee, *Nature Communications* **2024**, 15, 10498.
[35] Y. Zhao, F. Lan, Y. Li, Y. Huang, F. Wang, K. Wang, R. Zhang, *Advanced Materials* **2025**, 37, 2505012.
[36] S. Cho, T. Chang, T. Yu, S. L. Gong, C. H. Lee, *Science Advances* **2024**, 10, eadk4295.
[37] W. Yang, S. Lin, W. Gong, R. Lin, C. Jiang, X. Yang, Y. Hu, J. Wang, X. Xiao, K. Li, Y. Li, Q. Zhang, J. S. Ho, Y. Liu, C. Hou, H. Wang, *Science* **2024**, 384, 74.
[38] T. Hou, C. Chen, R. Guo, S. He, Y. Zi, *Nature Communications* **2025**, 16, 11668.

[39] T. Li, Z. Li, W. Chen, J. Chen, H. Zhou, C. Wang, X. Cao, B. Li, S. Ding, L. Shi, *Nature Communications* **2025**, 16, 9023.
[40] J. Qu, Y. Tang, R. Zhou, T. Xiao, M. Lin, Z. He, Y. Song, J. Meng, L. Niu, F. Li, *Advanced Materials* **2026**, 38, e74209.